\documentclass[12pt]{article}

\usepackage{arxiv}

\usepackage[utf8]{inputenc} 
\usepackage[T1]{fontenc}    
\usepackage{hyperref}       
\usepackage{url}            
\usepackage{booktabs}       
\usepackage{amsfonts}       
\usepackage{nicefrac}       
\usepackage{microtype}      
\usepackage{setspace}       
\usepackage{lipsum}
\usepackage{graphicx}
\usepackage{subcaption}
\graphicspath{ {./images/} }
\usepackage{float}
\usepackage[australian]{babel}

\title{A Study on the Efficiency of the Indian Stock Market}

\author{
    Devansh Jain{$^1$} \\
    \And
    Manthan Patel$^2$ \\
    \And
    Aman Narsaria$^3$ \\
    \And
    Siddharth Malik$^4$ \\
    \And\\
    Department of Economics and Finance \\
    Birla Institute of Technology and Science, Pilani \\
    \texttt{ \{f20180798$^1$, f20180823$^2$, f20180743$^3$, f20180838$^4$\} @ pilani.bits-pilani.ac.in}
}

\begin{document}
\large
\maketitle
\begin{abstract}
The efficiency of the stock market has a significant impact on the potential return on investment. An efficient market eliminates the possibility of arbitrage and unexploited profit opportunities. This study analyzes the weak form efficiency of the Indian Stock market based on the two major Indian stock exchanges, viz., BSE and NSE. The daily closing values of Sensex and Nifty indices for the period from April 2010 to March 2019 are used to perform the Runs test, the Autocorrelation test, and the Autoregression test. The study confirms that the Indian Stock market is weak form inefficient and can thus be outperformed.
\end{abstract}


\section{Introduction}
Capital markets bring those who hold capital and those seeking capital together and provide a place where entities can exchange securities, such as equities and bonds. A capital market is efficient if all the information available about assets is already incorporated into the assets' price. An efficient market thus eliminates the possibility of arbitrage and unexploited profit opportunities. 

The Efficient Market Hypothesis, in the general form, states that asset prices reflect all available information. The hypothesis is based on the idea of the Random Walk Theory. The theory proclaims that stocks take a random and unpredictable path, making all methods of predicting stock prices futile in the long run. It considers technical analysis undependable because it is based on established trends, and fundamental analysis undependable due to the often-poor quality of information collected and its ability to be misinterpreted. Thus, Random Walk Theory believes it is impossible to outperform the market without assuming additional risk. The Efficient Market Hypothesis also disregards both technical and fundamental analysis since it believes that the present price is entirely independent of past price and cannot predict future changes.

The term "Efficient Market" was introduced by Fama in 1965. In 1970, Fama introduced a classification of the hypothesis based on the level of information reflected in stock prices.

\subsection{Weak Form EMH}
Weak form states that today's stock prices reflect all the data of past prices. The data includes trading information derived from market data such as past prices and trading volume. Thus, future price changes can only be the result of new information becoming available. 
Weak form thus implies that technical analysis is undependable and useless. However, it does allow the use of fundamental analysis to identify undervalued and overvalued stocks.

\subsection{Semi Strong Form EMH}
Semi strong form states that stock prices reflect all the publicly available information and adjust quickly to absorb new public information. Public information includes past price information, macroeconomic factors, company's annual reports, announcements, among other information. 
Semi strong form thus implies that neither technical analysis nor fundamental analysis can lead to high returns since the information gained through such analysis is already publicly available and thus incorporated into stock prices. However, it does allow the use of private information to gain an advantage in trading.

\subsection{Strong Form EMH}
Strong form states that all information, both the information available to the public and any information not publicly known, is wholly accounted for in current stock prices. There does not exist any information that can give an investor an advantage on the market.
Strong form thus implies that investors cannot make returns on investments that exceed average market returns, regardless of information retrieved or research conducted.

The present study aims to analyze the Weak Form Market Efficiency in the Indian stock market based on the theory of the Efficient Market Hypothesis. The researchers take into account data for ten years (2010 - 2019) to conclude if both NSE (National Stock Exchange) and BSE (Bombay Stock Exchange) are weak form efficient or not.

\section{Literature Review}
\paragraph{Kashif Hamid et al., 2010 \cite{article}}
This study aims at testing the weak form market efficiency of the stock market returns of Asian countries and Australia. Monthly observations are taken from January 2004 to December 2009. Mathematical techniques like Autocorrelation, Ljung-Box Q-statistic Test, Runs Test, Unit Root Test, and the Variance Ratio are used to investigate whether the selected equity markets follow the Random Walk Model at an individual level or not. The study concluded that the monthly prices do not follow random walks in all the 14 countries of the Asian-Pacific region. Hence, investors can stream of benefits through the arbitrage process from profitable opportunities across these markets.

\paragraph{A. Q. Khan et al., 2011 \cite{khan2011testing}}
This study tested the weak form of market efficiency on the Indian Capital Market based on indices of two major stock exchanges of India. The efficiency of the Indian capital market is analyzed using the daily closing values of the indices of NSE and BSE by employing Runs Test, a non-parametric test. The study concludes that the Indian capital market neither follows a random walk model nor is weak form efficient.

\paragraph{Sachin et al., 2014 \cite{sachin2014efficiency}}
This study investigated the Indian stock market's efficiency using twenty-three stocks in different sectors of the National Stock Exchange. They tried to examine the random walk hypothesis using the serial autocorrelation test (a non-parametric test) using daily stock prices. The results show that the Indian stock market is weak form inefficient.

\paragraph{Hartika Arora et al., 2017 \cite{arora2017testing}}
This paper attempts to verify the weak form of efficient market hypothesis using interval return data for the top 10 frequently traded stocks and the Nifty 50 from 1st January 2009 to 31st March 2011. The author tests the weak form of efficiency using high-frequency data, which is required to capture the stock market's intraday predictability characteristics. Mathematical analysis has been done with the Augmented Dickey-Fuller (ADF) test to check the data's stationarity, ARMA model to verify autocorrelations, and GARCH (1,1) model for symbolizing volatility. The results of these statistical models' present evidence for the nonexistence of the weak form of efficiency, thereby providing an opportunity to investors towards exploiting the predictable characteristics of the market
through trading.

\paragraph{Matteo Rossi et al., 2018 \cite{rossi2018efficient}}
This research studies some of the most critical market irregularities in France, Germany, Italy and Spain stock exchange indexes between 2001-2010. In this study, to mathematically verify the distribution of the returns and their autocorrelation, statistical methods such as the GARCH model and the OLS regression are used. This analysis does not show significant anomalies but some market-specific effects.

\paragraph{Ajju Patel et al., 2018 \cite{patel2018testing}}
This paper investigated the weak form of market efficiency. Three-year daily closing points were taken from the Bombay Stock Exchange (BSE) official website, from 1st April 2015 to 31st March 2018. The Runs test was used to analyze data. The study concluded that the market is not efficient enough to adjust readily by the news regarding the factors that may affect the stock market prices, so there may be chances where investors who are in touch with factors may outperform the market.

\paragraph{Vidya A, 2018 \cite{vidya2018empirical}}
This research examined the concept of Market Efficiency on ten securities using runs test and autocorrelation test. It is found that the relationship between the past stock price of sample companies and their future stock price is very meager. This shows that price changes are random, and the market is efficient in the weak form.

\section{Data}
The data used for the study consists of daily closing values for Sensex (Bombay Stock Exchange) and Nifty (National Stock Exchange) indices for the period from April 2010 to March 2019.

\section{Methodology}
The weak form of EMH is mainly motivated from the theory of Random Walk. As a result, the tests aim to check if successive price changes were independent of each other. \\
In this study, Runs Test, Autocorrelation and Autoregression are used to test the weak form efficiency of the stock markets. The tests have been performed using Microsoft Excel, STATA and Python.

\subsection{Runs Test}
The runs test is a non-parametric test that depends only on the sign of the price changes and not the magnitude of the price. The test is used to decide if a data set is from a random process or not. The test only takes into consideration the direction of changes in the time series. A run is defined as a series of increasing values or a series of decreasing values.\\ The hypotheses for the Runs test are defined as: \\
\textbf{Null Hypothesis $H_0$}: The data set is from a random process. \\
\textbf{Alternate Hypothesis $H_1$}: The data set is not from a random process.

The $Z$ distribution used to test the mentioned hypotheses for the Runs test is given by:
\begin{equation}
    Z = {\frac {R - \mu} {\sigma}}
\end{equation}
$R$ = Number of runs \\
$\mu$ = Mean of runs \\
$\sigma$ = Standard deviation of runs

\subsection{Autocorrelation Function ACF(k) \cite{awasthi2015indian}}
The autocorrelation function ACF(k) for the time series $Y_{t}$ and the k-lagged series $Y_{t-k}$ is defined as:
\begin{equation}
    ACF(k) = {\frac {\sum _{t=1-k}^{n} (y_{t} - \bar{y}) (y_{t-k} - \bar{y})} {\sum _{t=1}^{n} (y_{t} - \bar{y})^2}
    }
\end{equation}
Here, $\bar{y}$ is the overall mean of the concerned series with $n$ observations.

The standard error of ACF(k) is given by:
\begin{equation}
    se_{ACF(k)} = {\frac {1} {\sqrt{n - k}}}
\end{equation}

For a sufficiently large $n$ ($n \ge 50$), the standard error of ACF(k) is approximated by:
\begin{equation}
    se_{ACF(k)} = {\frac {1} {\sqrt{n}}}
\end{equation}

The hypotheses for the ACF test are defined as: \\
\textbf{Null Hypothesis $H_0$}: Correlation does not exist. \\
\textbf{Alternate Hypothesis $H_1$}: Correlation exists.

The $t$ distribution used to test the mentioned hypotheses for ACF(k) is given by:
\begin{equation}
    t = {\frac {ACF(k)} {se_{ACF(k)}}}
\end{equation}

\subsection{Autoregression}
Autoregressive models operate under the premise that past values have an effect on current values. In a way, we can define autoregression as multiple regression having past values as its independent variables, similar to predictors used in simple multiple regression.

The model used for autoregression is given by:
\begin{equation}
    Index_t = Constant + \sum _{p}^{t-1} \beta_p (Index)_{t-p}
\end{equation}

\newpage
\section{Results}
\subsection{Runs Test}
The results for runs test for both Sensex and Nifty are shown in Table \ref{tab:runs}. It can be can seen that the calculated Z-statistic is less than 5\% critical value (-1.96). Thus, at 5\% significance, the null hypothesis is rejected. This implies that the data set is not random. Hence, the runs test proves that the present prices depend on past prices. 

\begin{table}[H]
    \caption{Runs Test}
    \centering
    \begin{tabular}{|l|r|r|}
    \toprule
    Statistic     & Sensex & Nifty  \\
    \midrule
    Number of Runs & 1037 & 1033\\
    Number of Positive Runs & 1158 & 1153 \\
    Number of Negative Runs & 1053 & 1047 \\
    Mean of Runs & 1104.1 & 1047 \\
    Standard Deviation of Runs & 23.45 & 23.39 \\
    Calculated Z-Statistic & -2.857 & -2.768 \\
    \bottomrule
    \end{tabular}
    \label{tab:runs}
\end{table}

\subsection{Autocorrelation Test}
\begin{figure}[H] 
    \centering
    \begin{subfigure}[H]{0.45\textwidth}
        \includegraphics[width=\textwidth]{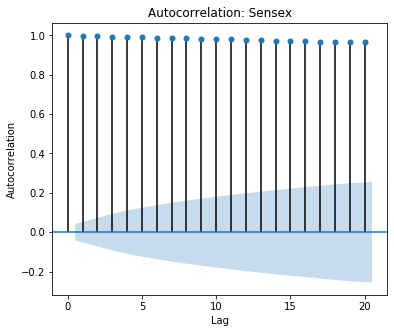}
        \caption{Autocorrelation: Sensex}
        \label{fig:acf_s}
    \end{subfigure}
    \hfill
    \begin{subfigure}[H]{0.45\textwidth}
        \includegraphics[width=\textwidth]{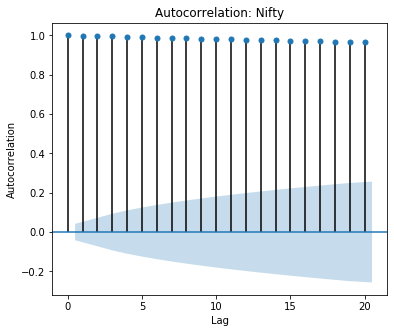}
        \caption{Autocorrelation: Nifty}
        \label{fig:acf_n}
    \end{subfigure}
    \caption{Correlation Graphs}
    \label{fig:corr}
\end{figure}

The correlation graph for Sensex (Figure \ref{fig:acf_s}) and Nifty (Figure \ref{fig:acf_n}) are shown in Figure \ref{fig:corr}. The length of each spike in the graph indicates the value of autocorrelation. The spikes for the first 20 lags are above 0.95 for both, Sensex and Nifty. Hence, autocorrelation exists and present prices depend on the past prices.

The results for autocorrelation test for both Sensex and Nifty are shown in Table \ref{tab:acf}. The autocorrelation value of every lag for both Sensex and Nifty is approximately fifty times of the standard error and so, the calculated t-statistic will be 50. Thus, the null hypothesis is rejected at 5\% significance level. This implies that autocorrelation exists. Hence, the autocorrelation test proves that the present value is correlated with past values.

\begin{table}[H]
    \caption{Autocorrelation Test}
    \centering
    \begin{tabular}{| c | c | c |}
    \toprule
    Lags & Nifty & Sensex \\
    \midrule
    1 & 0.998 & 0.998 \\
    2 & 0.996 & 0.996 \\
    3 & 0.994 & 0.994 \\ 
    4 & 0.993 & 0.992 \\
    5 & 0.991 & 0.990 \\
    6 & 0.989 & 0.988 \\
    7 & 0.987 & 0.986 \\ 
    8 & 0.985 & 0.985 \\ 
    9 & 0.984 & 0.983 \\
    10 & 0.982 & 0.981 \\
    11 & 0.980 & 0.979 \\
    12 & 0.978 & 0.977 \\ 
    13 & 0.977 & 0.975 \\
    14 & 0.975 & 0.974 \\
    15 & 0.973 & 0.972 \\
    16 & 0.972 & 0.970 \\
    17 & 0.970 & 0.969 \\
    18 & 0.969 & 0.967 \\
    19 & 0.967 & 0.966 \\
    20 & 0.966 & 0.964 \\
    Standard Error & 0.021 & 0.021 \\
    \bottomrule
    \end{tabular}
    \label{tab:acf}
\end{table}

\subsection{Autoregression}
\paragraph{Sensex} The results for autoregression for Sensex are shown in Table \ref{tab:reg_s}. The coefficients for Lag 1 and Lag 2 are significant as p-value is less than 0.05. Thus, the null hypothesis is rejected at 5\% significance level.  This implies that $\beta_p \ne 0$. Hence, autoregression proves that the present value depends on past valeus for Sensex.

\begin{table}[H]
    \caption{Autoregression: Sensex}
    \centering
    \begin{tabular}{c c c c}
    \toprule
    Regression Statistics \\
    \midrule
    R Squared & 0.968 \\ 
    Adjusted R Squared & 0.967 \\
    \toprule
    & Coefficients & t-value & p-value \\
    \midrule
    Intercept & 7.907 & 0.42 & 0.6790 \\
    Lag 1 & 1.077 & 50.60 & 0.0000 \\
    Lag 2 & -0.077 & -3.61 & 0.0003 \\
    \bottomrule
    \end{tabular}
    \label{tab:reg_s}
\end{table}

\paragraph{Nifty} The results for autoregression for Nifty are shown in Table \ref{tab:reg_n}. The coefficients for Lag 1 and Lag 2 are significant as p-value is less than 0.05. Thus, the null hypothesis is rejected at 5\% significance level.  This implies that $\beta_p \ne 0$. Hence, autoregression proves that the present value depends on past valeus for Nifty.

\begin{table}[H]
    \caption{Autoregression: Nifty}
    \centering
    \begin{tabular}{c c c c}
    \toprule
    Regression Statistics \\
    \midrule
    R Squared & 0.999 \\ 
    Adjusted R Squared & 0.999 \\
    \toprule
    & Coefficients & t-value & p-value \\
    \midrule
    Intercept & 2.902 & 0.50 & 0.6142 \\
    Lag 1 & 1.077 & 50.56 & 0.0000 \\
    Lag 2 & -0.077 & -3.61 & 0.0003 \\
    \bottomrule
    \end{tabular}
    \label{tab:reg_n}
\end{table}

\section{Conclusion}
Financial market efficiency is an essential issue for investors, researchers, analysts, and regulators of emerging markets like India. This study examines the weak form efficiency of the Indian Stock market based on the two major stock exchanges of India, viz., BSE (Bombay Stock Exchange) and NSE (National Stock Exchange). Various statistical techniques are employed, viz Runs Test, Autocorrelation Test and Autoregression Test. The Runs Test rejects the presence of a random walk and supports that the Indian market is not weak form market efficient. It implies that the Random Walk Model cannot determine the movement of the stock market indices. The Autocorrelation test confirms that a correlation exists between the past values and present values. The Autoregression test shows that present values do, in fact, depend on past values. 

The practical implication of inefficiency in stock markets is that it may lead to variation in the expected returns of stocks. In the state of inefficiency, stock prices may not reflect the fair value of the stocks. As a result, the companies with a lower fair value of shares may find it challenging to raise capital. Moreover, it also shows that the market is not efficient enough to adjust readily by the news regarding the factors that may affect the stock market prices. Thus, investors who are in touch with such factors may outperform the market.

\bibliographystyle{unsrt} 
\bibliography{paper}
\end{document}